\documentclass[twocolumn,showpacs,prc,superscriptaddress]{revtex4}
\usepackage{amsmath,amssymb}
\providecommand{\LyX}{L\kern-.1667em\lower.25em\hbox{Y}\kern-.125emX\@}

\usepackage[T1]{fontenc}
\usepackage[latin1]{inputenc}
\usepackage{graphics}
\usepackage{bm,hyphenat,xspace}
\usepackage{ulem}
\makeatother

\begin{document}

\newcommand {\mbf}[1]{{\mathbf{#1}}}
\newcommand {\vecg}[1]{\mbox{\boldmath{$#1$}} }
\newcommand{\be}{\begin{eqnarray}}
\newcommand{\ee}{\end{eqnarray}}
\topmargin -1cm

\title{Probing nucleon-nucleon interactions in 
 breakup of a one-neutron halo $^{11}$Be on a proton target}

\author{E.~Cravo}
\affiliation{Centro de F\'{\i}sica Nuclear, Universidade de Lisboa,
Av.\ Prof.\ Gama Pinto 2, 1649-003 Lisboa, Portugal}
\author{R.~Crespo}
\affiliation{Centro de F\'{\i}sica Nuclear, Universidade de Lisboa,
Av.\ Prof.\ Gama Pinto 2, 1649-003 Lisboa, Portugal}
\affiliation{Departamento de F\'{\i}sica, Instituto Superior T\'ecnico, 
Taguspark, Av. Prof. Cavaco  Silva,  Taguspark, 2780-990 Porto Salvo, 
Oeiras, Portugal}
\author{A.M.~Moro}
\affiliation{Departamento de F\'{\i}sica Atómica, Molecular e Nuclear,
Universidad Seville, Universidad de Sevilla, Apartado 1065, 41080 Sevilla, 
Spain}
\author{A.~Deltuva}
\affiliation{Centro de F\'{\i}sica Nuclear, Universidade de Lisboa,
Av.\ Prof.\ Gama Pinto 2, 1649-003 Lisboa, Portugal}
\email{raquel.crespo@tagus.ist.utl.pt//edgar@cii.fc.ul.pt//moro@us.es
//deltuva@cii.fc.ul.pt//}

\date{\today}

\begin{abstract}
A comparison between full few-body Faddeev/Alt-Grassberger-Sandhas (Faddeev/AGS)
and continuum-discretized coupled channels (CDCC)
calculations is made for the resonant and nonresonant
breakup of $^{11}$Be on proton target
at  63.7 MeV/u incident energy. A simplified two-body
model is used for  $^{11}$Be  which involves an inert $^{10}$Be(0$^+$) core
and a valence neutron. 
The sensitivity of the calculated observables to the nucleon-nucleon potential
dynamical input is analysed. We show that
with the present NN and N-core dynamics the results remain a puzzle
for the few-body problem of scattering from light exotic halo nuclei.

\end{abstract}

\maketitle

Halo nuclei are few-body light nuclear systems that appear at the neutron 
drip line. The study of these nuclei
is providing new theoretical challenges to
the nuclear reaction theory, a key tool to interpret and 
extract reliable nuclear structure information 
from the new generation of  very precise experimental
measurements. 

When describing the scattering of stable nuclei from halo nuclei
it is crucial to handle its few-body character. In addition it is necessary
to treat  all opening channels (elastic, inelastic, transfer
and breakup) on equal footing. Yet, it is fair to say that a tight control
on the underlying physics of the reaction mechanisms has not been achieved.

We aim to shed light on the relevant dynamical aspects of 
the reaction framework.

Recently  the resonant and non-resonant breakup resulting from the scattering of 
$^{11}$Be from a proton target at 63.7 MeV/u was measured at MSU
in order to obtain information on the $^{11}$Be continuum by nuclear 
excitation \cite{Shivastava04}. 
The experimental results were compared 
with the calculated observables using the 
Faddeev/Alt-Grassberger-Sandhas (Faddeev/AGS) scattering framework
\cite{faddeev60,Alt,Glockle} and making use of a simplified two-body
model for  $^{11}$Be in terms of an inert $^{10}$Be(0$^+$) core and a 
valence neutron \cite{Cravo09}. 
Due to the experimental energy resolution, the 
breakup angular distribution results contained integrated  contributions
from  ranges of relative neutron-core energy.
It was found in \cite{Cravo09} that
in the case where the relative core-neutron energy is 
integrated around the resonance 
$E_{\rm r} =1.275$ MeV in the energy range $E_{\rm rel} = 0-2.5$ MeV
the Faddeev/AGS approach  reproduced fairly well the 
shape distribution of the data although underestimating the maximum value 
of the breakup observable.  
A large discrepancy was found between the predictions 
of the Faddeev/AGS approach and those made by the
continuum-discretized coupled channels (CDCC) framework
done in the works of  \cite{Shivastava04,Summers}.
Given the previous $p+^{11}$Be benchmark calculations \cite{benchmark1}
one should expect some differences between Faddeev/AGS and CDCC results,
however, the different dynamic input may be responsible for the
disagreement between the predictions of \cite{Cravo09} and 
\cite{Shivastava04,Summers} as well.
In this work we aim to make a fair comparison between the two scattering 
frameworks using the same dynamical and structure inputs in order to establish
the source of the discrepancy and shed light on the sensitivity of
the observables to some key aspects of the dynamic input.
In any 3-body reaction approach to our working example 
of the breakup of $^{11}$Be by a proton target, 
one needs as input the 3-pair 
potentials: nucleon-nucleon (NN), N-core and p-core. In the late nineties
high precision NN potentials wich reproduce the NN data below 350 MeV lab energy
with $\chi^2$/datum $\sim 1$ were developed such as the CD Bonn 
\cite{CDBONN} and AV18 \cite{AV18} potentials. 
However, puzzling  3 body problems
remain (such as the Ay discrepancy) where the current NN potentials are unable
to reproduce the data. It is relevant to know to what extent our three-body
observables are sensitive to this underlying dynamical input.

The Faddeev/AGS is a non-relativistic three-body  multiple scattering 
framework that can be used to calculate all relevant three-body observables. 
We use  here the odd man out notation
for the three  interacting particles (1,2,3)
which means, for example, that the potential between
the pair (2,3) is denoted as $v_1$.
According to this reaction framework, one needs to evaluate the
operators $U^{\beta \alpha}$, whose on-shell matrix elements are 
the transition amplitudes. These operators  are obtained by solving the three-body AGS  
integral equation~\cite{Alt,Glockle} 
\be
U^{\beta \alpha} = \bar{\delta}_{\beta \alpha} G_0^{-1}
+ \sum_{\gamma} \bar{\delta}_{\beta \gamma }
t_\gamma G_0 U^{\gamma \alpha} \; ,
\label{Uba2}
\ee
with $\alpha, \beta, \gamma = (1,2,3)$, ($\beta = 0$ in the final 
breakup state).
Here,  $ \bar{\delta}_{\beta \alpha} = 1 - {\delta}_{\beta \alpha}$ and 
the pair transition operator is
\be
 t_{\gamma} =  v_{\gamma} +  v_{\gamma} G_0  t_{\gamma} \; ,
\ee 
where $ G_0 $ is the free resolvent
$G_0 = (E+i0 - H_0)^{-1}$,
and $E$ is the total energy of the three-particle system in the center of mass 
(c.m.) frame.

\begin{figure*}
\resizebox*{0.50\textwidth}{!}
{\includegraphics{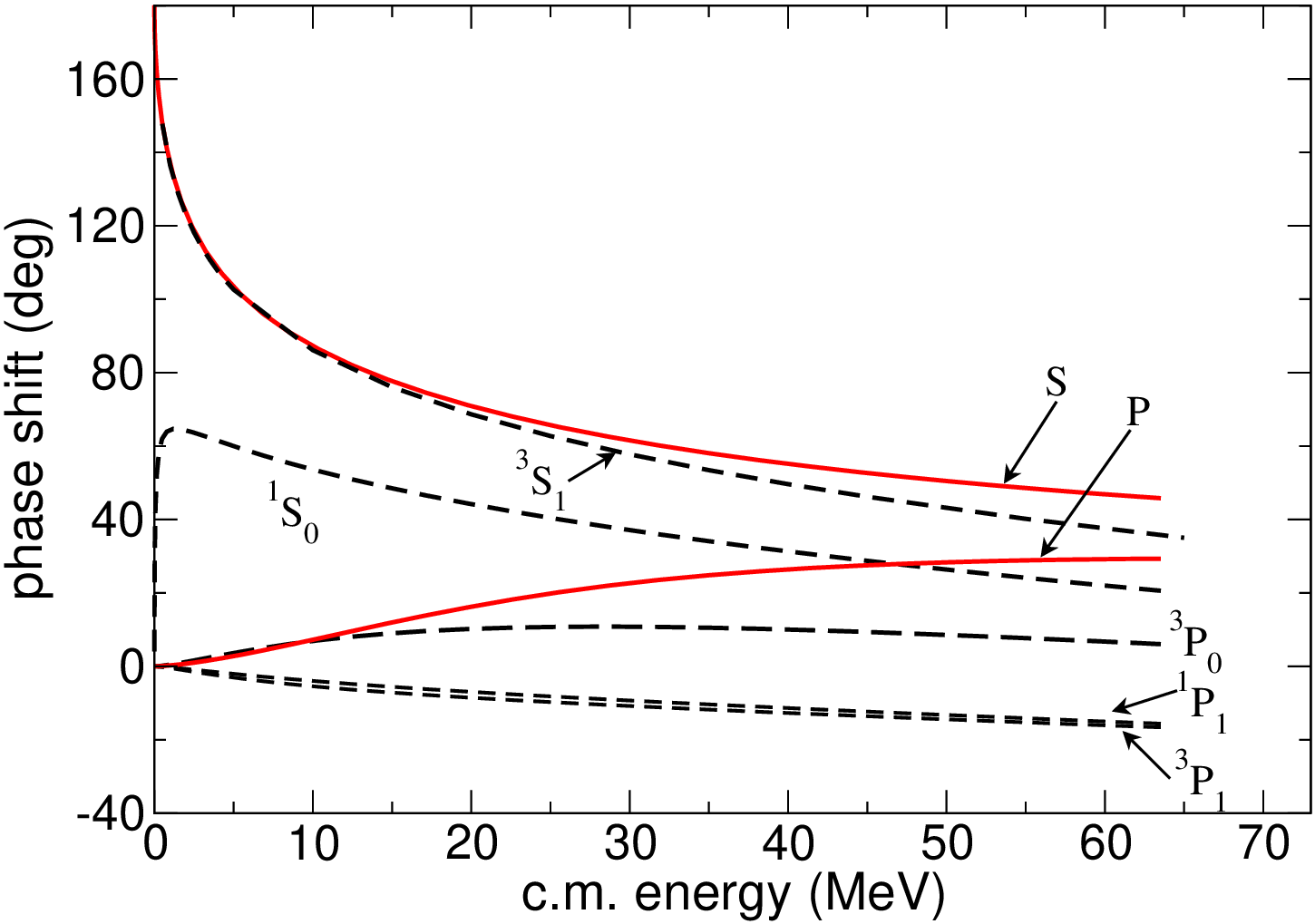}} 
 \caption{ \label{Fig:phase-shifts}(Color online) NN phase shifts  for S and P
    partial waves obtained from the CDBonn potential (dashed lines) 
    and Gaussian potential (solid lines) as indicated by the labels, 
    where the depth and range parameters were adjusted
    to fit the deuteron binding energy, and denoted  in the text 
    as $G_{TE}$ (light lines)\cite{gaussian}.
}
\end{figure*}

In our calculations Eq.~(\ref{Uba2}) is solved exactly 
in momentum space after partial wave decomposition 
and discretization of all momentum variables. The Pad\'e 
method~\cite{Pade} is used to sum the multiple scattering series. 
We include the nuclear interaction between all three pairs,
and the Coulomb interaction between the proton and $^{10}$Be, 
following  the technical developments implemented in 
Refs.~\cite{Del05b,Del06b} and the breakup observables are calculated as 
summarized in detail in \cite{Crespo08b}.

\begin{figure*}
\resizebox*{0.35\textwidth}{!}
{\includegraphics{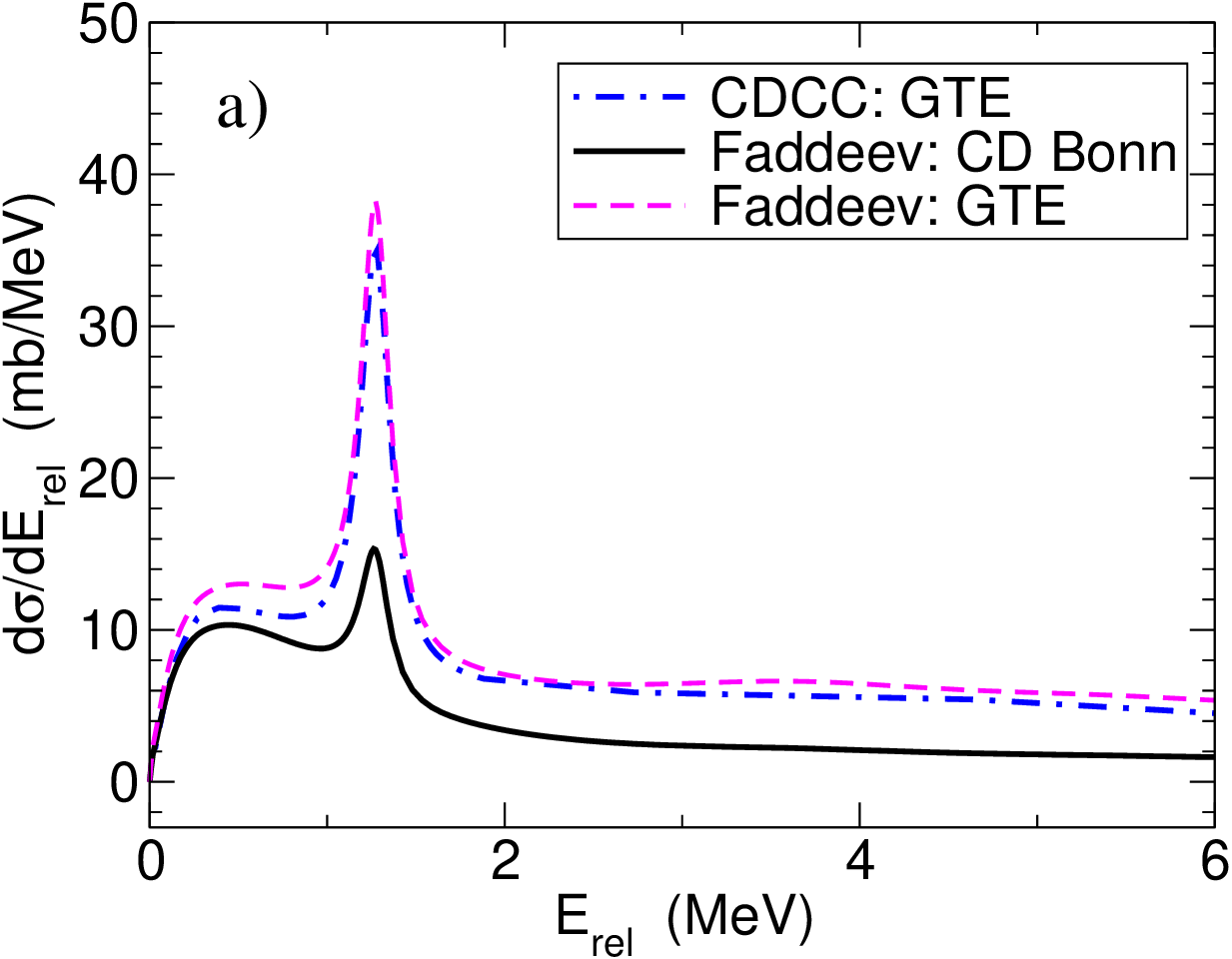}} 
\resizebox*{0.35\textwidth}{!}
{\includegraphics{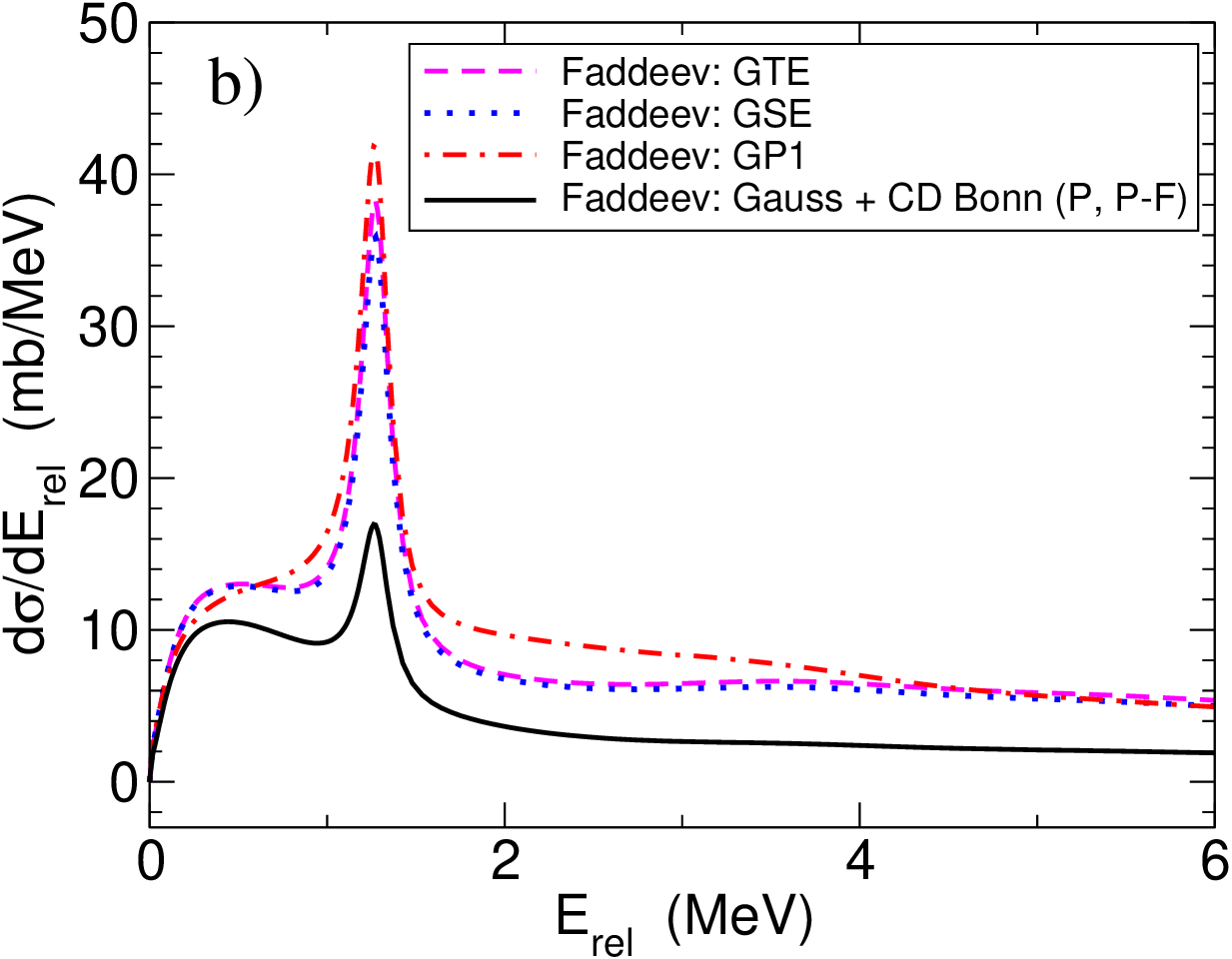}}
\caption{\label{Fig:espectrum}
 (Color online) Energy spectrum for 
  p($^{11}$Be,p)$^{10}$Be n at 63.7~MeV/u integrated over 
the whole angular range. The NN interactions are described in the text.
}
\end{figure*}

The 3-body Continuum-Discretized Coupled-Channels (CDCC) \cite{CDCC,CDCC0}
reaction framework consists in solving the Schr\"odinger equation 
in a model space in which the three-body wavefunction is expanded 
in the internal states (bound and continuum resonant and non-resonant states) 
of the two-body projectile $H_p$. 
 In practical calculations, the continuum spectrum  has 
to be truncated in excitation energy, and discretized into a set of 
square-integrable functions. The most widely used 
discretization method is the called \textit{binning} method, in which 
the continuum is divided into a set of energy 
intervals; for each interval, or \textit{bin}, a representative function, 
$\phi_{\alpha}(\mbf{r})$, is constructed by superposition of the 
 \textit{true} scattering states within the bin interval. 
The total three-body wavefunction is expanded in terms of these 
representative functions as
\begin{equation}
\Psi^{\rm CDCC}_{\mbf{K}_0}(\mbf{R}, \mbf{r}) = \sum_{\alpha=0}^N 
\phi_{\alpha}(\mbf{r})\omega_{\alpha}(\mbf{R})
\label{CDCCwf}
\end{equation}
where in here $\alpha$=0 refers to the projectile ground state,
$\mbf{K}_0$ is  the  incident wave number of the projectile in the c.m. frame,
$\mbf{R}$ the relative distance between the c.m. of
the projectile and the target and $\mbf{r}$ the relative distance between 
the valence particle and the core. 
The  wave functions $\omega_{\alpha}(\mbf{R})$ of the projectile-target 
relative motion are solutions of the coupled-channel equations
\be
\left( E_{\alpha}-T_R-V_{\alpha\alpha}(\mbf{R})\right) \omega_{\alpha}(\mbf{R})
= \sum_{\beta \neq \alpha}V_{\alpha\beta}(\mbf{R})\omega_{\beta}(\mbf{R})~,
\label{ceq}
\ee
where $E_{\alpha}=E-\varepsilon_{\alpha}$ and the coupling potentials are
\be
V_{\alpha\beta}(\mbf{R}) = \langle \phi_{\alpha}
|  \sum_{j=C,v}V_{jt}(\mbf{R}, \mbf{r}) | \phi_{\beta} \rangle~.
\ee
 The CDCC method was originally developed  in order to
incorporate the effect of the breakup channels in 
deuteron induced reactions. The  proton-target and neutron-target 
interactions used in these calculations are 
usually taken as optical potentials adjusted to reproduce the elastic 
scattering at the same energy per nucleon, i.e., 
$E_p\approx E_n \approx E_d/2$. For 
nucleon-nucleus scattering these optical potentials are in many cases 
well represented by 
a simple, local, {\it L}-independent interaction comprising a central and, 
maybe, a spin-orbit term \cite{ohkubo}.  
This has been also the standard choice in the application of the CDCC 
method to the scattering of other 
weakly bound nuclei ($^{6,7}$Li, $^{11}$Be, $^8$B, etc) by medium-mass or 
heavy targets. However,  this simple 
prescription might not be appropriate to describe the scattering of halo nuclei 
on protons, because in this 
case one of the fragment-target interactions is the NN potential, 
which is known to be strongly {\it L}-dependent. Furthermore, the 
absence of an imaginary (absorptive) part in the NN interaction makes 
less clear the formal justification of the CDCC method as an accurate 
approximation of a three-body scattering problem \cite{austern}. 

 In this context, the calculations presented in this work arise from 
a twofold motivation. Firstly, we aim to study the 
importance of using a realistic NN interaction in the description of the 
scattering of one-neutron halo nuclei by a proton target. 
In addition, by comparing the CDCC with the 
Faddeev/AGS calculations using the same three-body Hamiltonian 
we  check to what extent the CDCC method provides  an accurate reaction 
framework to the solution of a three-body scattering problem, 
following previous work done in ref. \cite{benchmark1}. 

In order to study the sensitivity of the calculated observables to 
the underlying NN potential, we have compared the  Faddeev/AGS 
calculations using the realistic   
CD Bonn  \cite{CDBONN} and AV18 \cite{AV18} NN potentials with those 
obtained making use of a simple {\it L}-independent potential.
For the latter, we first consider a simple
Gaussian potential $v_\mathrm{pn}(r)=-v_0 \exp[-(r/r_0)^2]$, 
with $v_0=72.15$~MeV and $r_0=1.484$ fm. These parameters are 
 adjusted to fit the deuteron binding energy and low-energy $^3S_1$ 
proton-neutron phase-shifts. This parametrization, here denoted as $G_{TE}$, 
has been used in several works to generate the deuteron states 
in $d$+A CDCC calculations  (see e.g.\ Ref.~\cite{gaussian}).  
The scattering phase-shifts associated to this potential for S- and P-waves 
are compared in Fig.~\ref{Fig:phase-shifts} with those 
obtained with the realistic  CD Bonn potential. 
It is seen that the $^3S_1$ phase-shifts are well described by the 
Gaussian parametrization up to a c.m.\ energy of $\approx 20$ MeV, 
but they are strikingly different from the realistic phase-shifts 
for the singlet S-wave and the P-waves.

In addition,  we have performed some additional 
Faddeev/AGS calculations considering several {\it L}-dependent modified 
Gaussian potentials: the $G_{SE}$ ($G_{P1}$) where the S (P) partial waves were
modified taking the parameters from  a Gaussian potential adjusted
to reproduce  the singlet scattering length \cite{gaussian} and all the other
partial waves kept the same as the $G_{TE}$ potential.  
Finally, we also consider the $G_{P2}$ example case where the P and 
P-F partial waves were taken from the realistic CD Bonn potential 
and all the other partial waves kept the same as the $G_{TE}$ potential.  
The proton-core and neutron-core pair interactions are taken as described 
in Ref. \cite{Cravo09}.  

\begin{figure*}
\resizebox*{0.35\textwidth}{!}
{\includegraphics{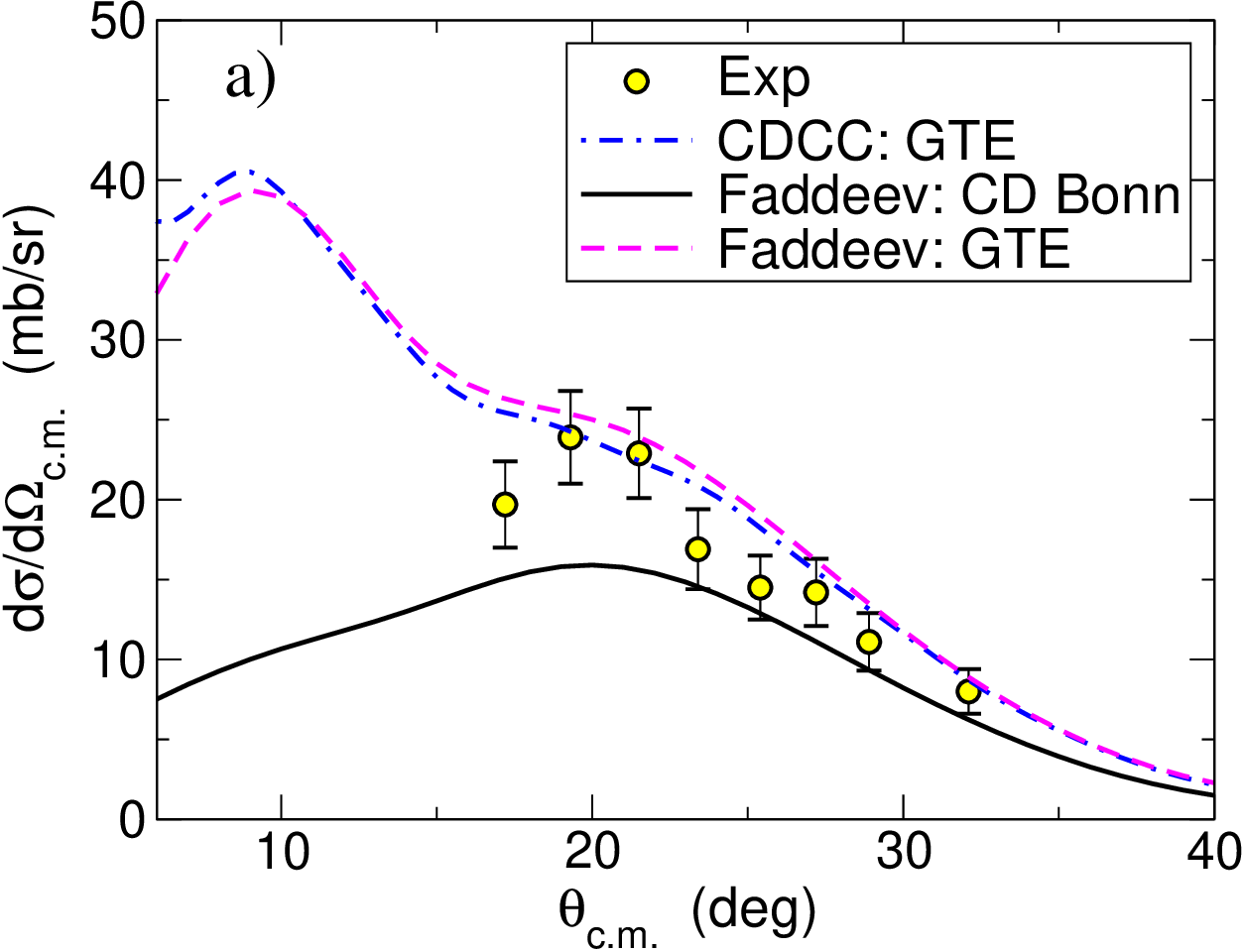}}
\resizebox*{0.35\textwidth}{!}
{\includegraphics{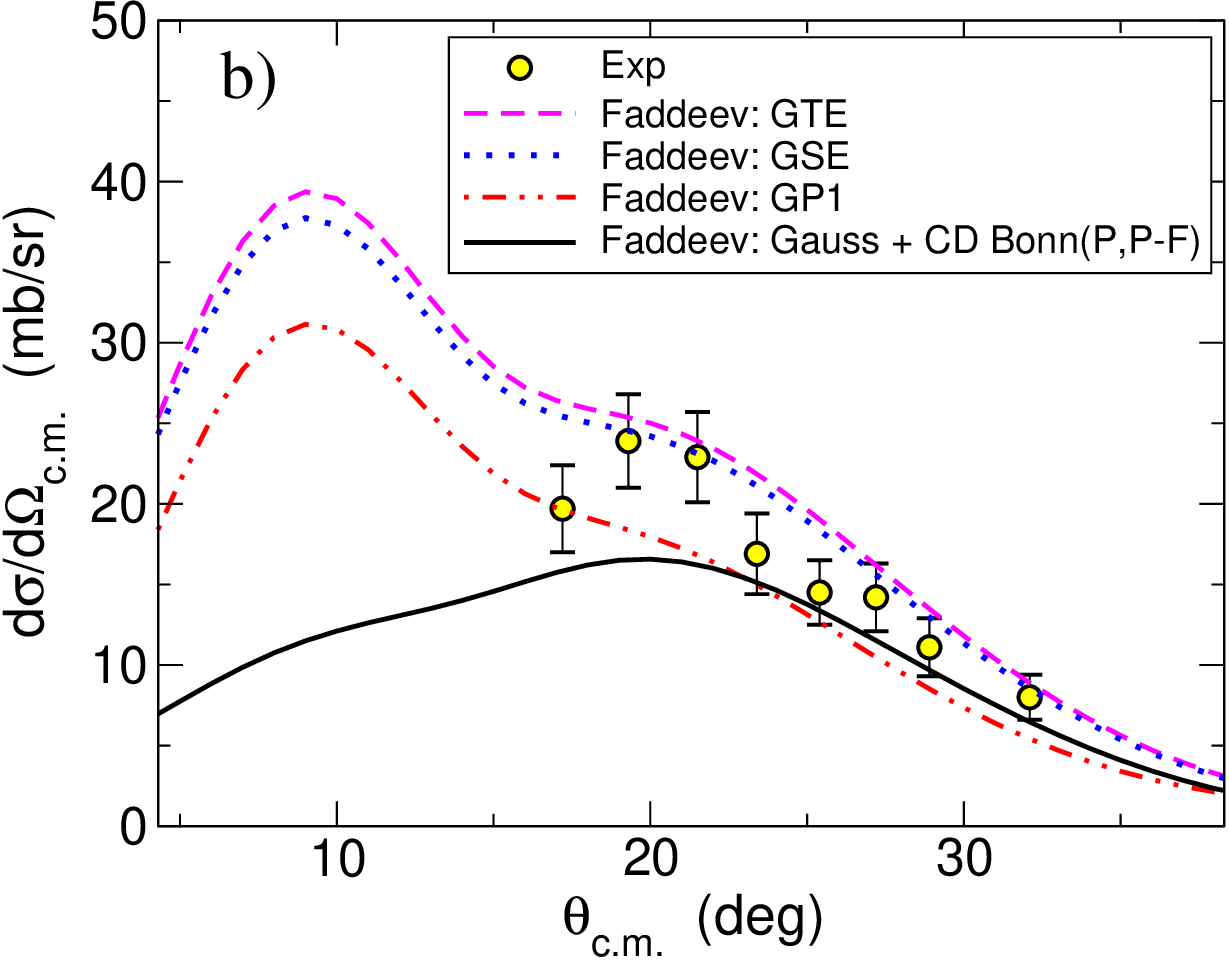}}
\par
\resizebox*{0.35\textwidth}{!}
{\includegraphics{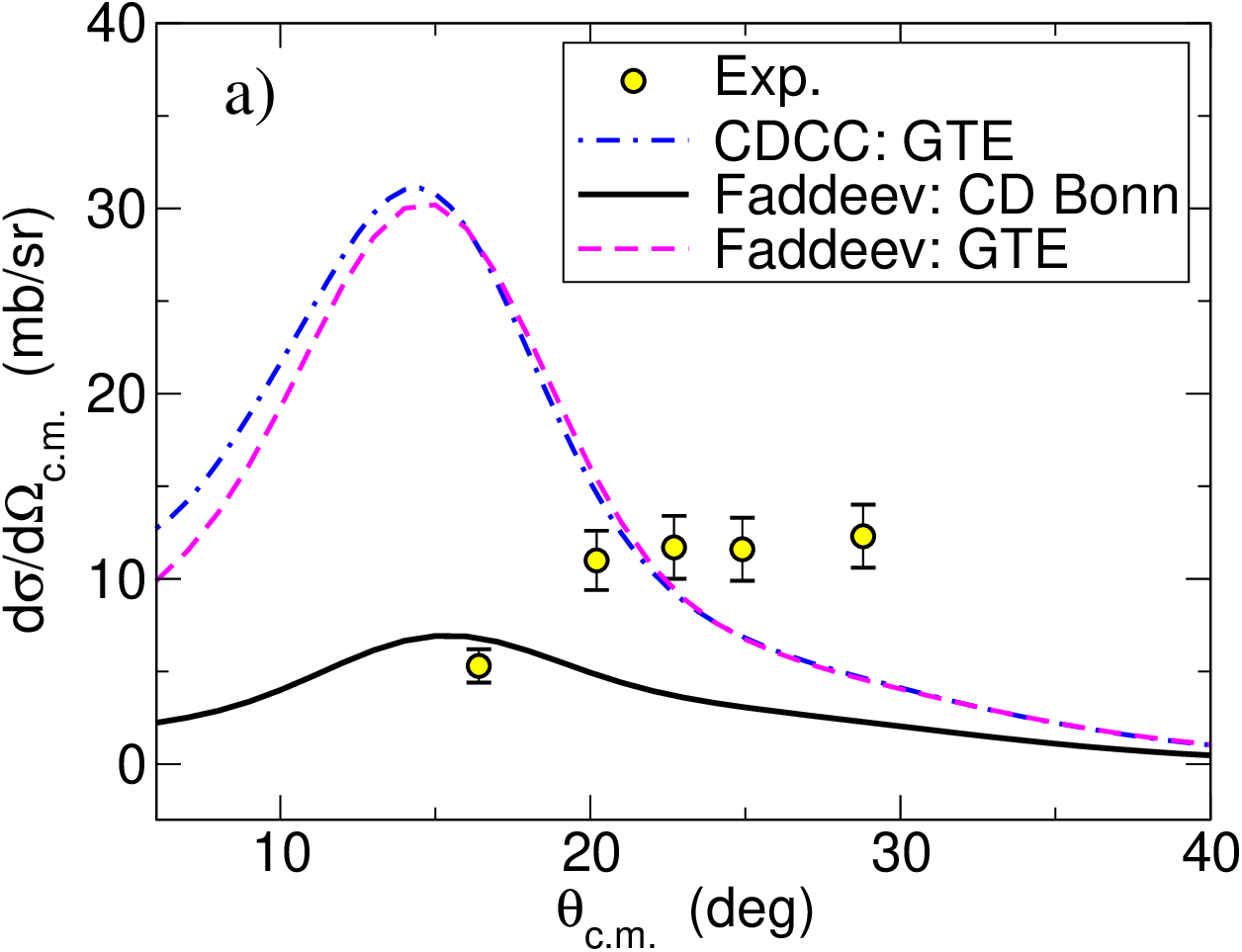}}
\resizebox*{0.35\textwidth}{!}
{\includegraphics{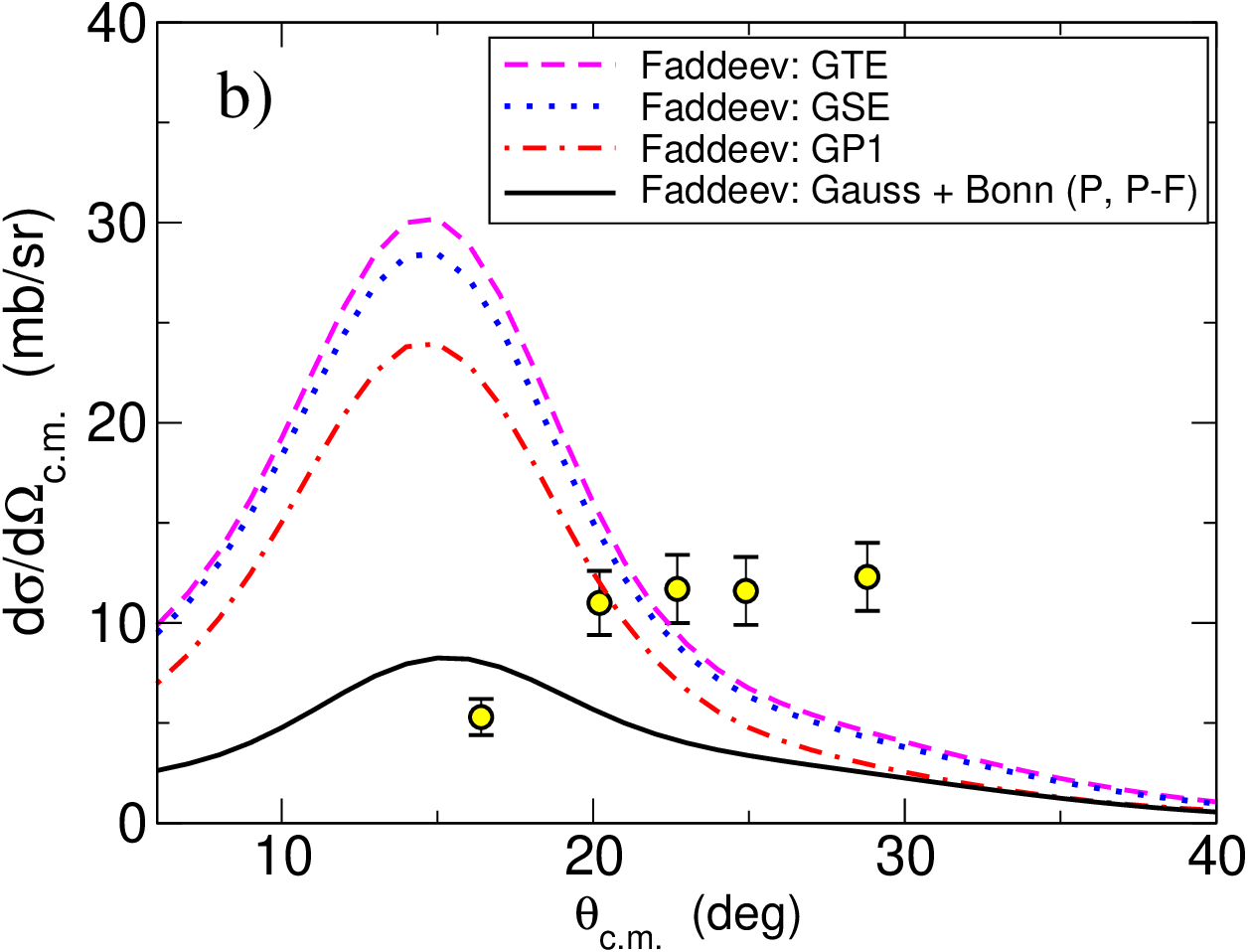}}
\caption{\label{Fig:angular}
 (Color online) Angular distribution for the breakup 
  p($^{11}$Be,p)$^{10}$Be n at 63.7~MeV/u   integrated over
the energy range $E_{\rm rel}=0-2.5$ MeV (upper part) and integrated over
the energy range $E_{\rm rel}=2.5-5.0$ MeV (lower part). The NN interactions
are described in the text. 
}
\end{figure*}

In the solution of the Faddeev/AGS equations we include
$n$-$p$ partial waves with relative orbital angular momentum $L \leq 6$, 
$p$-$^{10}$Be with  $L \leq 21$, and $n$-$^{10}$Be with $L \leq 6$.
Three-body total angular momentum is included up to 40.

As for the CDCC equations we  discretized  the n-$^{10}$Be continuum 
using energy bins
for partial waves $L \leq 4$, and up to a maximum excitation energy of 40 MeV, 
above  breakup threshold. The coupled equations were solved for
total angular momentum up to 60, and the solutions matched to their 
asymptotic form at a distance of 60 fm. Both Coulomb and nuclear couplings 
were included and treated on equal footing.  
For the energy spectrum 
the number of bins  was increased 
for excitation energies below 2 MeV in order to provide a finer description 
of the resonance region. 

In Fig.~\ref{Fig:espectrum} we represent the calculated energy spectrum
$d\sigma/dE_{\rm rel}$ that emerges by integrating 
the semi-inclusive cross section over the solid 
angle $d\Omega_{\rm c.m.}$. As follows from  Fig.~\ref{Fig:espectrum}a)
the Gaussian {\it L}-independent potential G$_{TE}$ (dashed line)
considerably overpredicts the non-resonant background 
and also the resonant peak with respect to the 
Faddeev/AGS result calculated with the CD Bonn potential (solid line).
The predicted results using CD Bonn are  indistinguishable
from those which make use of the AV18 \cite{AV18} potential and 
are not represented in the graph.
The CDCC result (dashed-dotted line) that  uses  the same G$_{TE}$ potential 
is very close to the corresponding Faddeev/AGS calculation except at the 
very low relative energies  in a region around 1 MeV, 
where the CDCC cross section is somewhat smaller.  
The sensitivity to the underlying NN low partial waves interactions 
is shown in Fig.~\ref{Fig:espectrum}b) which shows that this observable
is sensitive to a relistic treatment of the NN potential in particular of
the P-waves.

In Fig.~\ref{Fig:angular} we show the breakup angular distribution
$d\sigma/d\Omega_{\rm c.m.}$. We have not included very small angles
$\theta_{\rm c.m.} < 5^{\circ}$ where there are no data and
the convergence of the Faddeev/AGS results with respect to 
the Coulomb screening radius is slow.
Due to the energy resolution of the experimental setup,
the relative core-neutron energy is integrated around the resonance 
$E_{\rm r} =1.275$ MeV in the energy range $E_{\rm rel} = 0-2.5$ MeV 
in the upper part of the figure.
The angular distribution calculated with the
Faddeev/AGS approach and using the CD Bonn NN potential
(solid curve) shown in Fig.~\ref{Fig:angular}a)
describes  the overall shape distribution of the data reasonably well,
although clearly  underestimating
the data around $\theta_{c.m.}=20$ degrees by about 40$\%$. 
The dependence on the calculated observable on the proton-core optical potential
was studied. Using a potential that fits the elastic proton core data 
\cite{Summers}  increases the calculated observable at 
$\theta_{c.m.}=20$ degrees by less than  $10\%$. 
The origin of this disagreement needs to be further investigated.
The Faddeev/AGS  results are again 
indistinguishable when using the AV18 \cite{AV18} potential.
It follows that these 3 body observables are  insensitive to the
choice of the realistic NN potential and probe essentially the NN scattering
on-shell behaviour.

Upon substituting the realistic potential by a
$L$-independent Gaussian G$_{TE}$ potential the Faddeev/AGS results 
for the calculated angular distribution (dashed solid line) are
significantly enhanced at small
c.m. angles and overall become similar to the predictions
of the CDCC approach (dashed-dotted line). 
As seen in Fig.~\ref{Fig:angular}b),
in this case, the angular distribution is very sensitive to a 
realistic treatment of the
NN potential, in particular to the NN P-waves.

In the lower part of Fig.~\ref{Fig:angular} 
we show the breakup angular distribution
$d\sigma/d\Omega_{\rm c.m.}$, where the relative core-neutron energy is 
integrated over the energy range $E_{\rm rel}=2.5-5.0$ MeV. In this case
the Faddeev approach using the CD Bonn potential does not reproduce the
data.  
As in the upper case the Faddeev calculation using the GTE Gaussian 
interaction overestimates the realistic calculation and overall becomes similar 
do the CDCC approach. Again the nonresonant breakup at
higher relative energies  is also very sensitive to a realistic treatment
of the NN low partial waves.

Overall we can say that in our case study, there is a fairly good  
agreement between
the two microscopic reaction formalisms, the Faddeev/AGS and the CDCC,
 better than in  the benchmark calculation of
ref. \cite{benchmark1}. We attribute this difference to the fact that, 
in the referred work,  the breakup observables are studied 
with respect to the $^{10}$Be core (integrated with
respect to the neutron angle). That calculated inclusive breakup observable, 
was found to be dominated by configurations with small p-n angular momentum. 
These configurations are difficult to treat in a CDCC framework 
based in the expansion of the internal states 
of the n-$^{10}$Be [Eq.~\ref{CDCCwf}],  
and hence the slow convergence found in Ref.~[7] for those 
observables. In the present work, by contrast, we study selected 
regions dominated by small n-core relative energies and angular momenta,
for which the convergence of the CDCC approach is expected to be much faster. 
Similar conclusions
were achieved in the work of Ref. \cite{tostevin} for the breakup of
$^8$B on the $^{58}$Ni target.

In conclusion, we have performed full Faddeev-type and CDCC calculations for the
breakup of $^{11}$Be on a proton target at  63.7 MeV/u incident energy.
A simplified two-body model for  $^{11}$Be consisting
of an inert $^{10}$Be(0$^+$) core and a valence neutron has been used.
We have shown that the Faddeev results using a  $L$-independent $p$-$n$  
potential of Gaussian form (with the depth and range parameters adjusted
to fit the deuteron binding energy) considerably differ from those 
obtained with 
a realistic CD Bonn potential. The former are found to  overestimate the 
resonant and 
non-resonant breakup angular distributions calculated with the realistic 
NN interaction. 
 These results  strongly suggest  that, at least in this energy regime, 
these breakup observables 
are very sensitive to the underlying NN interaction, particularly to the 
P-wave.

We have also compared the Faddeev calculations with  CDCC 
calculations, using in both cases the simple NN {\it L}-independent 
Gaussian parametrization. The angular and energy breakup 
distributions are fairly similar, displaying only some small differences 
in the magnitude  of the breakup 
cross section at small excitation energies. The good  agreement between 
the two formalisms in this case study
is better than in  the previous benchmark calculation Ref. \cite{benchmark1}.

Since existing CDCC codes usually rely on simple central, 
{\it L}-independent fragment-target interactions, 
extensions of these codes, in order to incorporate realistic NN potentials 
in the calculation of the coupling potentials would be mandatory for 
future applications of the CDCC method in order to 
extract physically meaningful information from nucleon-nucleus scattering data.

We conclude that one can only extract reliable information from the breakup
of halo interaction if a realistic potential is used.
Different realistic NN potentials predict the same breakup observables
which then essentially  probe  the NN scattering
on-shell behaviour.
In addition,  the breakup angular distribution integrated around 
the resonance underestimates significantly the data by about 40$\%$ 
and clearly does not reproduce this observable when integration 
is made at a higher
relative core-neutron energy range. Further work should be performed 
to understand the physics of this discrepancy.
With the present NN and N-core dynamics these results remain a puzzle
for the few-body problem of scattering from light exotic halo nuclei.

{\bf Acknowledgements:}
The authors are supported by the FCT grant PTDC/FIS/65736/2006
and by the Complementary Action Portugal-Spain PORT2008-05.

\end{document}